\documentclass{article}
\usepackage{buckow}
\usepackage{amsfonts}
\usepackage{latexsym}
\def\ata{\mbox{arctanh}}

\def\IN{\mathbb{N}}
\def\IZ{\mathbb{Z}}

\def\unitmatrixDT%
 {\maththroughstyles{.45\ht0}\z@\displaystyle {\mathchar"006C}\displaystyle 1}
\begin {document}
\begin{flushright}
\small
VUB/TENA/00/10\\
MIT-CTP-3057\\
KUL-TF-00/30\\
{\tt hep-th/0101018}\\ 
December 14, 2000
\normalsize
\end{flushright}
\def\titleline{
D-branes and constant electro-magnetic backgrounds
}
\def\authors{
J.~Bogaerts\1ad, A.~Sevrin\1ad, J.~Troost\2ad, W.~Troost\3ad
and S.~van~der~Loo\1ad
}
\def\addresses{
\1ad Theoretische Natuurkunde\\
Vrije Universiteit Brussel\\
B-1050 Brussels, Belgium\\
\2ad Center for Theoretical Physics\\
MIT, Cambridge, MA 02139, USA\\
\3ad Institute for Theoretical Physics\\
K.U. Leuven, Belgium

}
\def\abstracttext{We probe the effective worldvolume theory of a set of 
coinciding D-branes by switching on constant electro-magnetic fields on 
them. The comparison of the mass spectrum predicted by string theory with 
the mass spectrum obtained from the effective action provides insights in
the structure of the effective theory. 
}
\large
\makefront
\section{Introduction}
The discovery of D-branes implies a novel way of looking at gauge 
theories. Indeed, the worldvolume degrees of freedom of a D$p$-brane are 
described by a $p+1$-dimensional field theory containing a $U(1)$ gauge field 
and $9-p$ scalar fields\footnote{Throughout the paper, we will ignore the 
fermionic degrees of freedom as they neither add to nor change our conclusions.}. 
The former describes an open string longitudinal to the 
brane while the latter describe the transversal fluctuations of the 
D$p$-brane. For slowly varying fields, the effective action is known to 
all orders in $\alpha '$: it is the ten-dimensional Born-Infeld action, 
dimensionally reduced to $p+1$ dimensions \cite{AT}, \cite{FT}. 

The situation becomes more interesting when several, say $n$, D$p$-branes 
are present. The mass of a string stretching between two branes is 
proportional to the shortest distance between these two branes. 
Ignoring the transversal coordinates, we have, as long as the branes are 
well separated, $n$ massless vector fields forming a $(U(1))^n$ gauge 
multiplet. However, once the branes coincide, $n(n-1)$ additional massless 
vector fields appear which correspond to oriented open strings connecting 
different branes. This enhances the gauge symmetry from $(U(1))^n$ to  
$U(n)$ \cite{EW1}. One expects the effective action to be some non-abelian 
generalization of the Born-Infeld action. However, as the notion of an 
acceleration term is ambiguous in a non-abelian theory, 
\begin{eqnarray}
D_i D_j F_{kl}=\frac 1 2 \{D_i , D_j\}F_{kl}-\frac i 2 {[}F_{ij}, F_{kl}
{]},
\end{eqnarray}
the concept of a slowly varying field is ambiguous too.
Nonetheless, we do have some information about the $U(n)$ non-abelian Born-Infeld 
action (NBI). Since it arises from the calculation of gluon scattering 
amplitudes in string theory, only one overall trace of the $U(n)$ matrices should be taken.
When switching off the off-diagonal modes of the gauge fields, it should 
reduce to the sum of $n$ copies of the abelian Born-Infeld action. 
Finally, the NBI was explicitely calculated through order $F^4$
\cite{GW}, \cite{BP}. Based on this and assuming that 
all terms proportional to anti-symmetrized products 
of fieldstrengths should be viewed as acceleration terms which in the
limit of slowly varying fields are ignored, a proposal was formulated 
for the NBI \cite{AT2}.
The action assumes a form similar to the abelian case but, upon 
expanding it 
in powers of the fieldstrength, one first 
symmetrizes over all fieldstrengths and subsequently one performs 
the group theoretical trace. Alternative possibilities are discussed 
in \cite{AN}. 

In the present paper, we review and extend some of the results obtained in
\cite{HT}, \cite{DST}. In those papers, the mass spectrum in the presence of constant
magnetic background fields was calculated
from the effective action and compared to predictions from string theory.
As will be demonstrated later on, this shows that the symmetrized trace 
proposal is flawed from order $F^6$ on.

A direct calculation of the effective action at higher order in $\alpha '$
would involve 
the analysis of at least a six-gluon scattering amplitude or the 
calculation of a five-loop $\beta$-function. As this does not seem 
feasible, different approaches are called for. 
One possibility, which we will review further in this paper, 
uses the mass spectrum as a guideline \cite{STT}. 
Another possibility uses $\kappa$-symmetry to fix the ordenings in the
effective action \cite{BDS}, \cite{here}. 

When finishing this paper, a preprint \cite{amb} appeared, 
where various aspects, 
mostly complementary to our work, of strings in constant 
electro-magnetic are studied. 

\section{Electro-magnetic backgrounds and Lorentz transformations}
Throughout the remainder of this paper, we mainly focus on D1- and
D2-branes. General results can be found in
\cite{DST}. In addition, we do not consider the transversal coordinates
as they provide no additional information.

We take two D2-branes wrapped once around a torus with cycles of length 
$L_1$ and $L_2$ and switch on magnetic fields
${\cal F}_{12}=b^{(0)}\sigma_0 + b^{(3)}\sigma_3$, where $\sigma_a$, $a\in\{1,2,3\}$ 
are the Pauli matrices and $\sigma_0$ is the $2\times 2$ unit matrix.
Flux quantization implies 
\begin{eqnarray}
b^{(0)}\pm b^{(3)}=\frac{2\pi}{L_1L_2}m_\pm, \quad m_\pm\in  \IZ.
\end{eqnarray}
Considering the Wess-Zumino term, which describes the coupling to the 
type  IIA RR background fields, one can show that this is equivalent to the 
statement that the branes have $m_+$ and $m_-$ D0-branes dissolved in 
them. We proceed by
choosing a gauge\footnote{
We are not very careful about the boundary conditions for the potentials on the torus. 
A detailed account of this can be found in e.g. \cite{wati2} or \cite{DST}.} 
such that ${\cal A}_1=0$ and ${\cal A}_2={\cal F}_{12} \, x^1$.
After T-dualizing in the 2 direction, we end up with two tilted D1-branes \cite{BDL}. 
The new transversal coordinate is given by \cite{wati1},
\begin{eqnarray}
X^2=2\pi\alpha '{\cal F}_{12}\,x^1 =
\left(  
\begin{array}{cc}
m_+ \hat L_2\frac{x^1}{L_1}&0\\
0& m_- \hat L_2\frac{x^1}{L_1}
\end{array}
\right),    \label{trc}
\end{eqnarray}
where $\hat L_2 =4\pi^2\alpha '/L_2$ is the length of the dual cycle. 
Eq. (\ref{trc}) clearly shows that the two D1-branes are wrapped once around
cycle~1 and $m_+$ and $m_-$ times resp. around cycle~2. The branes are rotated in
the 12 plane over angles $\arctan(2\pi\alpha '(b^{(0)}\pm
b^{(3)}))$. The angle 
$\phi$ between the two branes is given by 
\begin{eqnarray}
\phi&=& \arctan(2\pi\alpha '(b^{(0)}+b^{(3)}))-
\arctan(2\pi\alpha '(b^{(0)}-b^{(3)}))\nonumber\\
&=&\arctan\frac{4\pi\alpha 'b^{(3)}}
{1+(2\pi\alpha ')^2((b^{(0)})^2-(b^{(3)})^2)}.  \label{hoek}
\end{eqnarray}

We now turn to electric backgrounds. Consider a D1-brane 
wrapped around a circle in the 1 direction
and turn on a constant electric field along the brane of the form 
${\cal F}_{01}=e^{(0)}\sigma_0 + e^{(3)}\sigma_3$. In the gauge where 
${\cal A}_0=0$ and ${\cal A}_1={\cal F}_{01}x^0$, we end up, after
T-dualizing in the 1 direction, with two D0-branes boosted in the 1 direction. 
Their speeds $v_\pm$ 
and rapidities $\alpha _\pm$ are given by $v_\pm=2\pi\alpha '(e^{(0)}\pm e^{(3)})$ 
and $\alpha_\pm=\ata(2\pi\alpha ' 
(e^{(0)}\pm e^{(3)}))$ respectively. 
The fluxes are quantized, $\mbox{cosh}\alpha _\pm=m_\pm\in\IZ$, which 
gives the branes momentum $2\pi m_\pm /\hat L_1$ in the 1 direction. 
The relative velocity of the two branes is given by 
$\mbox{tanh}(\alpha _+-\alpha _-)= (v_+-v_-)/(1-v_+v_-)$.
Before T-duality, one can view an electric field with flux $m$ on 
a D1-brane as a bound state of a D1-brane with $m$ fundamental strings~\cite{EW1}. 

Finally, concerning the question of stability, as long as the 
electro-magnetic field has no component in the $\sigma_3$ direction, the 
resulting configuration is BPS with 16 supercharges preserved. Once a 
component in the $\sigma_3$ direction is turned on, one finds that for 
certain magnetic configurations, BPS states arise which preserve 2, 4, 6, 
or 8 supercharges. In the electric case, this never happens \cite{BDL}.

\section{The spectrum from Yang-Mills theory}
We consider a $U(2)$ Yang-Mills theory in the presence of a constant 
background ${\cal F}_{\mu \nu}={\cal F}^{(0)}_{\mu \nu} \sigma_0+
{\cal F}^{(3)}_{\mu \nu}\sigma_3$ with potential ${\cal A}_\mu $. 
After separating 
the potential in the sum of the background and the fluctuation, $A_\mu =
{\cal A}_\mu +\delta A_\mu $, one obtains for the part of the Lagrangian 
quadratic in the fluctuations,
\begin{eqnarray}
{\cal L}=-\frac 1 2 \left\{ \sum_{i=1}^2\partial_\mu \delta A_\nu^{(i)}
\partial^\mu \delta A^{(i)}{}^\nu+ 2{\cal D}^+_\mu \delta A_\nu^+
{\cal D}^-{}^\mu \delta A^-{}^\nu-8i{\cal F}_{\mu \nu}^{(3)}
\delta A^+{}^\mu \delta A^-{}^\nu\right\},  \label{kwad}
\end{eqnarray}
where we wrote,
\begin{eqnarray}
\delta A_\mu =
\left(  
\begin{array}{cc}
\delta A^{(1)}_\mu &\delta A^+_\mu \\
\delta A^-_\mu &\delta  A^{(2)}_\mu 
\end{array}
\right),
\end{eqnarray}
and ${\cal D}^\pm_\mu=\partial_\mu \mp 2i {\cal A}^{(3)}_\mu  $. 
We work in a background covariant Lorentz gauge.

Focussing on the magnetic case, ${\cal F}_{12}\neq 0$, we immediately read 
off the spectrum for the diagonal fluctuations,
\begin{eqnarray}
M^2=\left(\frac{2\pi m_1}{L_1}\right)^2+
\left(\frac{2\pi m_2}{L_2}\right)^2,\quad m_1,\, m_2\in \IZ.
\end{eqnarray}
The spectrum of the off-diagonal fluctuations is most easily obtained by 
passing to complex coordinates, $z=(x^1+ix^2)/\sqrt{2}$. The equations of 
motion of $\delta A^-$ become
\begin{eqnarray}
&&\left(\Box +2{\cal D}^-_{\bar z}{\cal D}^-_{z}-(2+4){\cal F}^{(3)}_{12}
\right)\delta A^-_z=0,\nonumber\\
&&\left(\Box +2{\cal D}^-_{\bar z}{\cal D}^-_{z}-(2-4){\cal F}^{(3)}_{12}
\right)\delta A^-_{\bar z}=0,
\end{eqnarray}
where we used that $[{\cal D}^-_z,{\cal D}^-_{\bar z}]=-2{\cal 
F}_{12}^{(3)}$ and $\Box$ is the d'Alambertian for the non-compact
directions. As shown in \cite{jt}, a complete set of eigenfunctions for 
the compact part of this operator in a gauge where $\partial_z 
{\cal A}_z=0$, is of the form 
\begin{eqnarray}
|m>= ({\cal D}^-_{\bar z})^m|0>,\quad m\in \IN,
\end{eqnarray}
where 
\begin{eqnarray}
|0>=e^{-2iz{\cal A}_z^{(3)}}\zeta (\bar z),
\end{eqnarray}
with $\zeta (\bar z)$ an anti-holomorphic function which satisfies appropriate 
boundary conditions on the torus. It is expressed in terms of 
$\theta$-functions \cite{jt}. From the fact that ${\cal D}^-_{ z} |0> 
=0$, one obtains the spectrum for the off-diagonal modes:
\begin{eqnarray}
M^2=2(2m+1\pm 2){\cal F}^{(3)}_{12},\quad m\in \IN.\label{YMmasses}
\end{eqnarray}

For an electric background field ${\cal F}_{01}$,
the situation is radically different. Instead of using complex 
coordinates, one uses light-cone coordinates $x^\pm=(x^1\pm x^0)/\sqrt{2}$.
The equations of motion for $\delta{\cal A}^-_\pm$ are now
\begin{eqnarray}
\left(\vec\nabla\cdot\vec\nabla  +2{\cal D}^-_-{\cal D}^-_++
2i(1\mp 2){\cal F}^{(3)}_{01}
\right)\delta A^-_\pm=0,
\end{eqnarray}
with $\vec\nabla$ the gradient in the transversal direction.
In order to diagonalize it, we expand the fluctuations in \cite{el},
\begin{eqnarray}
|m,y^\perp >= ({\cal D}^-_{-})^m e^{-2ix^+{\cal A}_+^{(3)}(x^-)}\zeta (x^-,y^\perp),
\quad m\in \IN,
\end{eqnarray}
where $y^\perp$ denote the transversal coordinates. This gives,
\begin{eqnarray}
\left(\vec\nabla\cdot\vec\nabla  +
2i(2m+1\mp 2){\cal F}^{(3)}_{01}
\right)|m,y^\perp >=0.
\end{eqnarray}
The imaginary part reflects the inherent instability of this system 
which manifests itself by Schwinger's 
pair production in an electric field \cite{schw0}, 
\cite{amb}, \cite{schw}.

\section{The spectrum from string theory}
The calculation of the mass 
spectrum for strings beginning and ending on a tilted brane is 
straightforward. Assume that the brane is rotated over an angle 
$\gamma$ into the 12 plane. The length of the D1-brane is $\tilde L_1= 
L_1/\cos\gamma$ and the length of a string attached to the brane and
wrapped once around the torus is $\tilde L_2= 
\cos\gamma \hat L_2$. With this we get the mass formula, 
\begin{eqnarray}
M^2&=&\left(\frac{2\pi m_1}{\tilde L_1}\right)^2 +
\left(\frac{m_2\tilde L_2}{2\pi\alpha '}\right)^2\nonumber\\
&=&\frac{1}{1+\tan^2\gamma }\left\{\left(\frac{2\pi m_1}{L_1}\right)^2+
\left(\frac{2\pi m_2}{L_2}\right)^2\right\},\quad m_1,\, m_2\in \IZ.
\end{eqnarray}
Using the results of previous section, we get $\tan\gamma=2\pi\alpha '
(b^{(0)}\pm b^{(3)})$.
In order to calculate the mass spectrum for strings stretching between the
two D1-branes at an angle $\phi$, we take one of them along the 1 axis and the other one 
rotated over an angle $\phi$ in the 12 plane. In this way, the boundary 
conditions become ,
\begin{eqnarray}
&&\left.\partial_\sigma X^1\right|_{\sigma=0}=
\left.\partial_\tau X^2\right|_{\sigma=0}=0,\nonumber\\
&&\left.(\partial_\sigma X^1+\tan \phi\, \partial_\sigma X^2)\right|_{\sigma=\pi}=
\left.(\partial_\tau X^2-\tan\phi\,\partial_\tau X^1)\right|_{\sigma=\pi}=0. \label{bdys}
\end{eqnarray}
Solving the equations of motion with these boundary conditions give
$Z(\sigma,\tau)=f(\sigma^+)+\bar f (\sigma^-)$, with 
$Z=(X^1+iX^2)/\sqrt{2}$, $\sigma^\pm=(\tau\pm\sigma)$ and
\begin{eqnarray}
f(\sigma^+)&=&i\sqrt{\frac{\alpha '}{2}}\sum_{m\in\IZ}
\frac{a^+_{m-\beta}}{m-\beta}e^{-i(m-\beta)\sigma^+}
\nonumber\\
\bar f(\sigma^-)&=&i\sqrt{\frac{\alpha '}{2}}\sum_{m\in\IZ}
\frac{a^-_{m+\beta}}{m+\beta}e^{-i(m+\beta)\sigma^-} ,
\end{eqnarray}
where $\beta\equiv\phi/\pi$. Upon quantizing this theory, we get 
low-lying states in the NS sector of the form
\begin{eqnarray}
|m>_\pm=\left(a^+_{-\beta}\right)^m \psi^\mp_{-\frac 1 2 \pm\beta}|0>_{NS}, \label{state}
\quad m\in\IN,
\end{eqnarray}
with mass \cite{BDL}, \cite{DST},
\begin{eqnarray}
M^2=(2\pi\alpha ')^{-1}(2m+1\pm 2)\phi.\label{StringMasses}
\end{eqnarray}
Note that the modes $\psi$ arise from the fermions in the NS sector.
The angle $\phi$ is given in eq. (\ref{hoek}) in terms of the magnetic 
fields. The appearance of a tachyon is a consequence of the fact that this 
configuration breaks all supersymmetry.

T-dualizing back along the 2 direction is equivalent to 
interchanging $\partial_\tau X^2 \leftrightarrow\partial_\sigma X^2$. The 
boundary conditions in eq. (\ref{bdys}) turn into those for a charged open string with 
no magnetic field at the $\sigma=0$ side and a magnetic field ${\cal 
F}_{12}= (2\pi\alpha ')^{-1}\tan\phi$ at the $\sigma=\pi$ side.

When calculating the spectrum for two D0-branes with relative rapidity 
$\alpha $, one solves the equations of motion combined with the boundary 
conditions
\begin{eqnarray}
&&\left.\partial_\sigma X^0\right|_{\sigma=0}=
\left.\partial_\tau X^1\right|_{\sigma=0}=0,\nonumber\\
&&\left.(\partial_\sigma X^0+\mbox{tanh} \alpha\, \partial_\sigma X^1)\right|_{\sigma=\pi}=
\left.(\partial_\tau X^1+\mbox{tanh}\alpha\,\partial_\tau X^0)\right|_{\sigma=\pi}=0.
\end{eqnarray}
Once this is done, one passes to the light-cone gauge and quantizes the 
system. In the case of D-branes with a relative velocity, the light-cone 
gauge becomes quite subtle. We refer to \cite{el} for more details. 

\section{Towards the effective action}
Comparing the spectrum obtained from an ordinary $U(2)$ Yang-Mills theory,
as in section 3, 
to that predicted by string theory, as in the previous section,
one obtains exact agreement only for 
small values of the background fields. In order to reproduce the string 
predictions correctly for arbitrary values of the backgrounds, one should 
use the Born-Infeld action instead of the Yang-Mills action. It is not 
hard to see that the diagonal fluctuations only probe the abelian part of 
the Born-Infeld action, 
\begin{eqnarray}
{\cal L}=-\sqrt{\det(\delta_\mu{}^\nu+2\pi\alpha 'F_\mu{}^\nu)}.
\end{eqnarray}
For a D2-brane with a constant magnetic background field 
${\cal F}_{12}$, the part quadratic in the fluctuations is given by
\begin{eqnarray}
{\cal L}=\frac 1 2 (\det {\cal G})^{-\frac{1}{4}}
\Big(  {\cal G}^{ij} \delta \tilde F_{0i} \delta \tilde F_{0j}
-\frac{1}{2} {\cal G}^{ij}  {\cal G}^{kl}
\delta\tilde F_{ik}  \delta\tilde F_{jl}  \Big),
\end{eqnarray}
where $\tilde F\equiv 2\pi\alpha ' F$ and
${\cal G}^{ij}=(1+\tilde{\cal F}^2)^{-1}\delta^{ij}$.  It is clear that 
this does reproduce the correct spectrum.

In \cite{DST}, the spectrum of the off-diagonal fluctuations
was calculated using the NBI action based on 
the symmetrized trace proposal. The result does not agree with the string 
calculation. This clearly showed that from order $F^6$ on, the symmetrized 
trace proposal for the NBI, is flawed.

Following an initial exploration of possible modifications at order 
$F^6$, \cite{bain}, a systematic investigation of the NBI through order $F^6$,
was performed in \cite{STT}. The mass spectrum is used as a guideline. 
We assume that the following properties hold for the $U(2)$ NBI:
\begin{itemize}
\item Only the fieldstrength should appear, not its derivatives.
\item The part quadratic in the fluctuations should reduce to eq. 
(\ref{kwad}) after performing a suitable coordinate transformation.
\end{itemize}
The first ansatz is a translation of our definition of slowly varying 
field configurations \cite{STT}. The second requirement is 
based on the observation that the spectrum is a rescaled Yang-Mills 
spectrum, compare eqs.~(\ref{YMmasses},\ref{StringMasses}).
In addition, because we use constant magnetic fields which are 
block diagonal in the Lorentz indices, only terms in the effective action
containing even numbers of Lorentz traced field strengths
contribute to our calculation \cite{STT}. 
Under these assumptions the $F^2$ and $F^4$ terms are fully 
determined. For the $F^6$ terms in the NBI, we find three different types:
\begin{itemize}
\item type 1: $tr(F_{\mu_1 \mu_2 } F^{\mu_2 \mu_3 } F_{\mu_3 \mu_4 }
F^{\mu_4 \mu_5 } F_{\mu_5 \mu_6 } F^{\mu_6 \mu_1 })$: 14 inequivalent 
ordenings,
\item type 2: $tr(F_{\mu_1 \mu_2 } F^{\mu_2 \mu_3 } F_{\mu_3 \mu_4 }
F^{\mu_4 \mu_1 } F_{\mu_5 \mu_6 } F^{\mu_6 \mu_5 })$: 9 inequivalent 
ordenings,
\item type 3: $tr(F_{\mu_1 \mu_2 } F^{\mu_2 \mu_1 } F_{\mu_3 \mu_4 }
F^{\mu_4 \mu_3 } F_{\mu_5 \mu_6 } F^{\mu_6 \mu_5 })$: 5 inequivalent 
ordenings.
\end{itemize}
Calculating the contribution of these terms to the mass spectrum under the 
assumptions outlined above and comparing it to the string calculation 
fixes 21 of the 28 parameters. Making the additional assumption 
that for D5-branes with static self-dual fieldstrength configurations the 
whole NBI collapses to the Yang-Mills action, \cite{brech}, fixes an 
additional 2 parameters.

It is not so surprising that the mass spectrum does not fix all parameters 
at order $F^6$. At this order one has for the first time terms of the form 
$tr([F,F][F,F][F,F])$, which for backgrounds living in the torus of $U(2)$ 
do not contribute to the spectrum.

\section{Conclusions}
The precise structure of the NBI remains an enigma. A direct calculation 
through gluon scattering amplitudes or $\beta$-functions looks
very hard. As 
we saw, just by using the mass spectrum, the $F^2$ and $F^4$ are easily 
and fully calculable. To a large extent, the $F^6$ term is determined as 
well. 

A different approach uses invariance under $\kappa$-symmetry as a way to 
determine the ordenings. It remains to be seen how much information this 
method will extract. However, it is encouraging that the 
$\kappa$-invariant Lagrangian deviates at low order from the symmetric 
trace proposal. We refer to elsewhere in this volume for more 
details \cite{here}. 

Perhaps one of the most interesting avenues is the study of non-abelian 
BPS states. These configurations should be solutions of the equations of 
motion. As explained in \cite{DST}, \cite{here}, some of these BPS configurations relate 
different orders of $F$ in the equations of motion. In such a way, one 
might hope for recursion relations to emerge.

In fact, a combination of both previous points might be considered.
Indeed, it would be interesting to investigate which non-linear
extensions of the Yang-Mills action are consistent with supersymmetry. In
9+1 dimensions, the supersymmetry algebra should be very restrictive, and
clues to the appropriate modifications of the susy-variations can be
obtained from the demand that the known BPS conditions, which translate
non-trivially in terms of background field strengths, solve the condition
for a trivial variation of the gluino.

Finally, a completely different approach would be to use some of the 
ideas in \cite{SW}. Consider the abelian case in the presence of a 
background. Instead of performing the calculations in this paper, one 
could just push the background field in the open string metric and a 
non-commutativity parameter. When calculating the mass spectrum
in the limit of slowly varying fields, the 
non-commutative Born-Infeld reduces to a $U(1)$ Yang-Mills theory with 
the open string metric instead of the flat one. Calculating the spectrum 
with this modified metric indeed reproduces the spectrum as predicted by 
string theory. As shown in \cite{SW}, one can reconstruct from this, order by order, 
the Born-Infeld action in the commutative limit. The generalization of 
these observations to the non-abelian case is quite subtle and presently 
under study \cite{BKS}.

\vskip0.5cm
\noindent
{\large \bf Acknowledgements}

\smallskip
\noindent
We like to thank E. Bergshoeff, J. de Boer, F. Denef, M. de Roo,
M. Henneaux, P. Koerber and P. van Baal for useful discussions. 
J.B., A.S., W.T. and S.v.d.L. are supported by the European Commission
RTN programme HPRN-CT-2000-00131, in which J.B., A.S. and S.v.d.L. are associated
to the university of Leuven.
The work of J.T. is supported in part by funds provided by the U.S. Department
of Energy under cooperative research agreement DE-FC02-94ER40818.


\end{document}